# Transport and electrical properties of cryogenic thermoelectric FeSb$_2$: the effect of isoelectronic and hole doping

Deepak Gujjar[a], Sunidhi Gujjar[b], V. K. Malik[c], and Hem C. Kandpal*[a]

[a] Department of Chemistry, Indian Institute of Technology Roorkee, Roorkee-247667, Uttarakhand, India

[b] Department of Chemistry, Banasthali Vidyapith, Tonk – 304022, Rajasthan, India

[c] Department of Physics, Indian Institute of Technology Roorkee, Roorkee-247667, Uttarakhand, India

* Corresponding author: hem.kandpal@cy.iitr.ac.in

Thermoelectric materials operating at cryogenic temperatures are in high demand for efficient cooling and power generation in applications ranging from superconductors to quantum computing. The narrow band-gap semiconductor FeSb$_2$, known for its colossal Seebeck coefficient, holds promise for such applications, provided its thermal conductivity value can be reduced. This study investigates the impact of isoelectronic substitution (Bi) and hole doping (Pb) at the Sb site on the transport properties of FeSb$_2$, with a particular focus on thermal conductivity (κ). Polycrystalline FeSb$_2$ powder, along with Bi- and Pb-doped samples, were synthesized using a simple co-precipitation approach, followed by thermal treatment in an H$_2$ atmosphere. XRD and SEM analysis confirms the formation of the desired phase pre- and post-consolidation using spark plasma sintering (SPS). The consolidation process resulted in a high compaction density and the formation of submicrometer-sized grains, as substantiated by electron backscattered diffraction

(EBSD) analysis. Substituting 1% of Bi and Pb at the Sb site successfully suppressed the thermal conductivity (κ) from ~15 W/m-K in pure $FeSb_2$ to ~10 and ~8.7 W/m-K, respectively. Importantly, resistivity measurements revealed a metal-to-insulator transition at around 6.5 K in undoped $FeSb_2$ and isoelectronically Bi-substituted $FeSb_2$, suggesting the existence of metallic surface states and provides valuable evidence for the perplexing topological behavior exhibited by $FeSb_2$.

## I. INTRODUCTION

The discovery of thermoelectricity, which enables the conversion of waste heat into electricity, has presented new opportunities for the research community to explore more efficient thermoelectric materials capable of operating across a wide temperature range.[1] This advancement is particularly significant as it addresses the ongoing energy crisis. However, the existing thermoelectric materials currently possess a relatively low conversion efficiency,[2] which can be quantified by their dimensionless figure of merit ($zT$) value at a specific temperature (T). The figure of merit is defined as $zT = \frac{S^2\sigma}{k}T$, where $S$ represents the Seebeck coefficient, $\sigma$ corresponds to the electrical conductivity, and κ denotes the total thermal conductivity which encompasses both electronic ($\kappa_e$) and lattice ($\kappa_L$) thermal conductivity.[2,3]

In the past few decades, remarkable advancements have been made in the field of thermoelectrics, focusing extensively on the discovery of materials that demonstrate excellent thermoelectric performance, especially at or above room temperature. Examples of such materials include $Bi_2Te_3$-based alloys, PbTe alloys, Si-Ge alloys, SnSe-based alloys, skutterudites, and half-Heusler alloys.[3-8] In 2021, Zhou et al. achieved the highest figure of merit of 3.1 at 783 K for polycrystalline SnSe.[9] However, there is still a lack of efficient thermoelectric materials (zT > 1) suitable for practical applications at cryogenic temperatures (T < 100 K).[10] Systems, such as $CeB_6$,[11] FeSi,[10,12] and $YbAl_3$,[12] have been extensively studied owing to their high Seebeck coefficient. In recent years, research has focused on $FeSb_2$ systems for cryogenic applications due to its record large Seebeck coefficient of 45000 μV/K at around 10 K temperature.[13-15]

$FeSb_2$ crystallizes into low-symmetry orthorhombic marcasite structure, where $FeSb_6$ distorted octahedra in *ab* plane, shares edge along the orthogonal to *ab* plane (i.e. *c*-axis) to form chains.[16] Theoretical studies identified $FeSb_2$ as a semiconductor with narrow band-gap ranging from 0.1-0.3 eV.[17] Several publications have considered $FeSb_2$ as *d*-electron based Kondo insulator (KI)[18,19] characterized by strongly correlated electrons or a small hybridization gap at the Fermi level. This gap is believed to be generated from the mixing of broad conduction bands with narrow *d*-bands[13] and found responsible for colossal Seebeck coefficient observed in $FeSb_2$ due to presence of large density of states (DOS) just below and above the gap.[14]

A record large value of Seebeck coefficient (*S*) of $FeSb_2$ contributes to its large power factor (*PF* = $S^2\sigma$) of 2300 μW / $K^2$, which is almost 40 times that of the best reported thermoelectric materials.[20,21] However, the excessively high thermal conductivity of the material significantly restricts its efficiency or figure-of-merit, rendering it unsuitable for practical applications. Therefore, researchers employed various methods such as introduction of defects, impurities, doping or grain boundaries for the suppression of lattice thermal conductivity to achieve a meaningful value of *zT*.[17,20] In 2012, Wang et al. separately substituted Cr and Co metal in place of Fe and Te metal in place of Sb of different concentrations in $FeSb_2$ and found 10 times enhancement in *zT* value from 0.005 to 0.05 in case of $Fe(Sb_{0.9}Te_{0.1})_2$ as compared to $FeSb_2$.[14] Koirala et al. introduced Cu nanoparticles in $FeSb_2$ nanostructured and achieved 90 % and 110 % enhancement in power factor and *zT*, respectively.[12]

In this study, we investigate the impact of substituting heavy metals Bi and introducing hole doping by substituting Pb in place of Sb on the transport properties of polycrystalline $FeSb_2$. Our primary focus is to understand the various scattering mechanisms that contribute to the reduction of thermal conductivity in heavy fermion systems. we report the effect of heavy metal Bi and Pb substitution on the transport properties of polycrystalline $FeSb_2$. Furthermore, we address the highly debated question regarding the presence of metallic surface states in the correlated *d*-electron system $FeSb_2$ below 10 K. In 2020, Xu et al. proposed the existence of metallic surface states in $FeSb_2$ single crystals at 6 K,[22] which we confirm in our polycrystalline $FeSb_2$ sample. Interestingly, these surface states persist in Bi substituted (isoelectronic with Sb) $FeSb_2$, but disappear in Pb substituted $FeSb_2$, thus paving the way for $FeSb_2$ as a topological insulator.

## 2. RESULTS AND DISCUSSION

*2.1. Structural and morphological analysis*

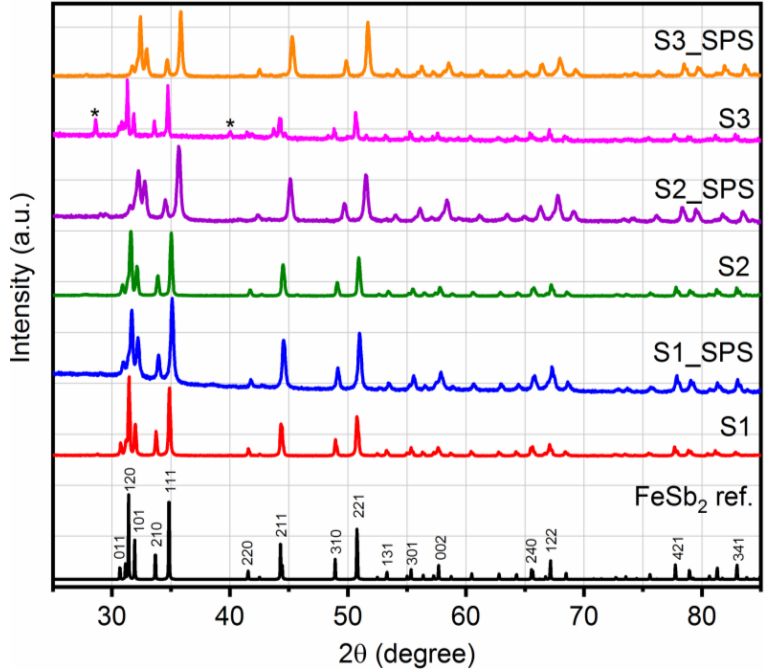

**Figure 1.** Shows the room temperature XRD patterns of all the represented FeSb$_2$ samples in the 2θ range of 20°– 90°. Sample S1 corresponds to pure FeSb$_2$, while sample S2 represents Bi substituted FeSb$_2$, FeSb$_{1.98}$Bi$_{0.02}$ and sample S3 denotes Pb substituted FeSb$_2$, FeSb$_{1.98}$Pb$_{0.02}$.

To initiate the substitution of Bi and Pb in place of Sb in FeSb$_2$, we synthesized a set of samples with varying concentrations (ranging from 1% to 5%) using an affordable co-precipitation method. However, for the subsequent investigations involving transport and electrical measurements, we focused exclusively on the samples substituted with 1% Bi (S2) and 1% Pb (S3). This decision was made because as the doping amount increased, an additional phase of Bi/Pb emerged, as illustrated in **Figure S1** (supplementary material). It is evident from the **Figure 1** that all the patterns exhibit almost similar characteristics and are consistent with the referenced XRD pattern (ICDD #: 04-010-4959) of orthorhombic structure with space group (SG) 58. All samples except Pb-substituted FeSb$_2$ powder (S3) were of pure phase, with S3 showing a minor metallic Sb phase indicated by an asterisk (*), which disappeared after spark plasma sintering (SPS) treatment. The same phenomenon of the appearance of excess Sb phase in powder form followed by its subsequent disappearance after Spark Plasma Sintering (SPS) treatment has been previously documented by Saleemi et al.[17] The powder X-ray diffraction (PXRD) patterns of all samples exhibit prominent reflection peaks at diffraction angles corresponding to the (120), (101), (210), (111), (211), and (221) crystallographic planes. Upon subjecting the samples to spark plasma sintering (SPS) treatment, the resulting microstructure displayed a preferred orientation along the (111) plane which was evident from the significant variations observed in the relative intensities of the aforementioned diffraction peaks. This observed preferred orientation along (111) plane has also been reported previously by Wang et al.[8] As expected, the substitution of larger-sized elements such as Bi and Pb resulted in peak broadening and a slight shift of the peaks towards lower diffraction angles, as evidenced by S2 and S3, respectively (**Figure 1**). However, it is essential to emphasize a notable observation of shifting of diffraction peaks towards higher angles after the SPS treatment. This shift occurred despite a slight increase in lattice parameters observed through Rietveld refinement (supplementary material, **Figure S2 & Table S1**). This intriguing behavior can likely be attributed to lattice strain and the presence of crystal defects,[23] which tend to be more pronounced in the substituted samples, as evidenced in the Williamson Hall plot (supplementary material, **Figure S3 & Table S2**).

The scanning electron microscopy (SEM) images and corresponding elemental mapping of the three samples after spark plasma sintering (SPS) treatment are shown in **Figure 2.** As depicted in **Figure S4** (supplementary material)**,** the particles of all three powder samples were found to have a non-uniform shape and size in the micrometer range. Following the SPS treatment, all the samples exhibited flat and immaculate surfaces at lower magnification, as shown in **Figures 2a, 2c, & 2e.** At higher magnification, nanograins embedded in the matrix were observed, indicating the occurrence of dynamic recrystallization during the sintering process, as displayed in **Figures 2b, 2d, & 2f.** The SEM images suggest that the $FeSb_2$ and $FeSb_{1.98}Bi_{0.02}$ samples have the same surface texture, as the nanograins generated due to recrystallization are individual grains in both samples. In contrast, the $FeSb_{1.98}Pb_{0.02}$ sample has a different surface texture, with nanograins agglomerated on the surface. The surface texture may play an important role in reducing thermal conductivity, as we will discuss in the transport properties section. The Energy Dispersive Spectroscopy (EDS) analysis (supplementary material, **Table S3**) after the SPS treatment, validates the successful replacement of Sb with Bi and Pb in accordance with the intended stoichiometric ratios.

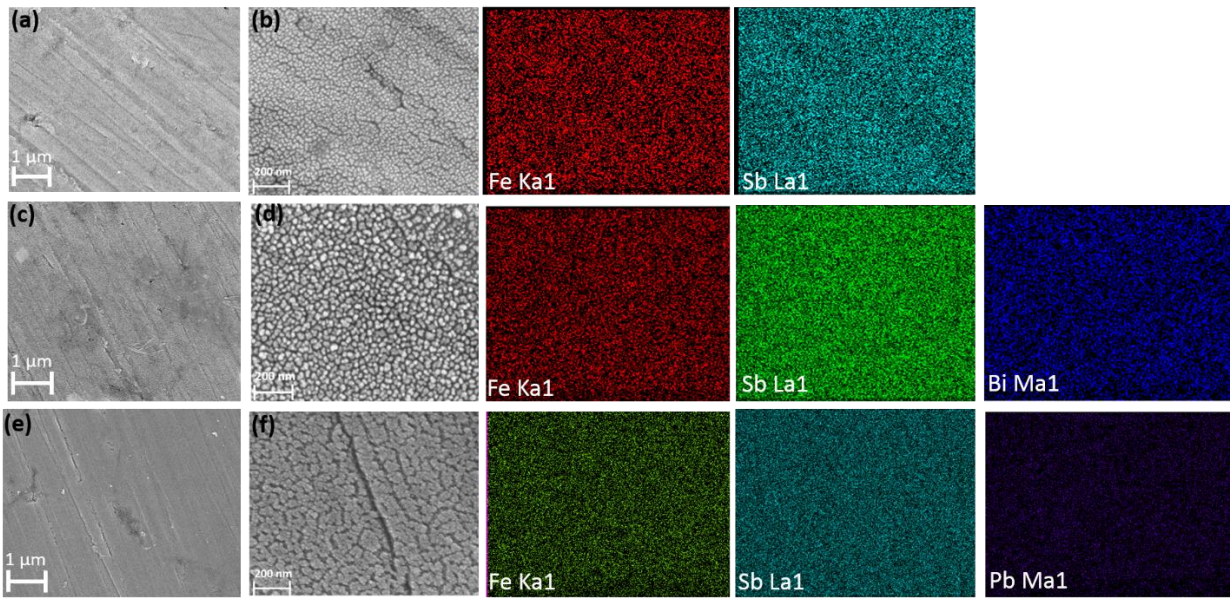

**Figure 2.** SEM images after spark plasma sintering treatment of pure $FeSb_2$ (a & b), $FeSb_{1.98}Bi_{0.02}$ (c & d), and $FeSb_{1.98}Pb_{0.02}$ (e & f) at two different magnifications and their corresponding EDS mapping.

*2.2. Transport and electronic properties*

To understand the effect of isoelectronic substitution and hole doping on the thermoelectric efficiency of FeSb$_2$, transport properties such as Seebeck coefficient, electrical resistivity, power factor and thermal conductivity, were measured for all three samples in the temperature range of 2-300 K & the obtained results were shown in **Figure 3.** The Seebeck coefficient (*S*) is negative up to about 135 K and becomes positive at high temperatures above 135 K, indicating the dominance of holes as charge carriers after 135 K in all three samples. The behavior of curve is in good agreement with the previously reported works.[8,14,16] The absolute value of Seebeck increases rapidly below 100 K and reaches its peak value at around 40 K, denoted as |$S_{max}$|. In the case of pure FeSb$_2$, the maximum absolute Seebeck coefficient is 245 µV/K, while FeSb$_{1.98}$Bi$_{0.02}$ and FeSb$_{1.98}$Pb$_{0.02}$ samples achieved values of 216 µV/K and 169 µV/K, respectively. It is important to acknowledge that the thermopower value |$S_{max}$|, as well as the temperature at which it exhibits maxima (T$_{max}$), is highly dependent on the synthetic conditions and methods employed during material synthesis. Several studies have reported a wide range of Seebeck coefficient values, spanning from 40 to 250 µV/K, along with variations in T$_{max}$, ranging from 10 to 70 K.[14, 16] The reduction in |$S_{max}$| associated with Bi substitution is likely a consequence of an augmented charge carrier concentration or a diminished Hall resistance, as evidenced in **Figure 5a**. Since the substitution of Bi for Sb involves isoelectronic chemical replacement, the incorporation of additional charge carriers can be attributed to the presence of crystal defects.[16] Conversely, the lowest recorded |$S_{max}$| value in the Pb-doped sample can be ascribed to the introduction of additional charge carriers of opposite polarity, i.e., holes. This introduction of holes in the crystal structure subsequently leads to a decrease in the absolute value of the Seebeck coefficient.

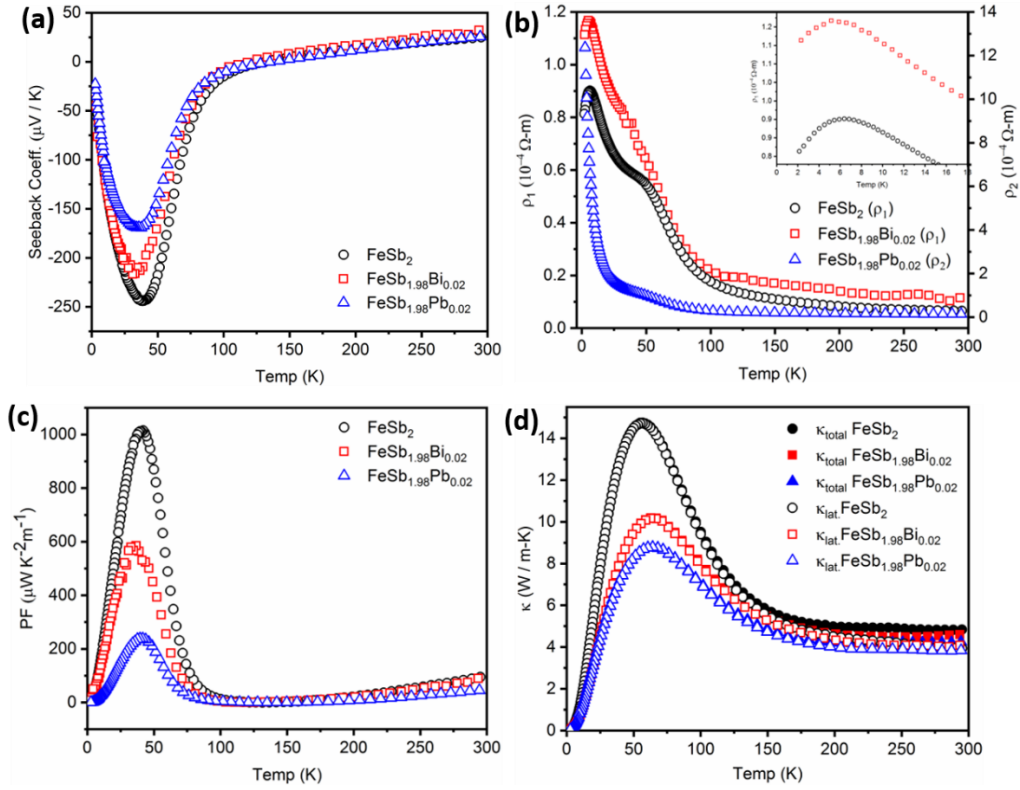

**Figure 3.** Temperature dependences of transport properties of $FeSb_2$ & Bi/Pb-substituted $FeSb_2$, (a) Seebeck coefficient, (b) electrical resistivity (inset represents the enlarged graph in the temperature range of 0-18 K), (c) power factor, and (d) thermal conductivity.

The electrical resistivity of pure $FeSb_2$ exhibits a metal-like temperature dependence up to a temperature of at least 7 K (**Figure 3b**). However, at higher temperatures, it demonstrates semiconductor-like behavior, with a decrease in resistivity from $9\times10^{-5}$ to $5\times10^{-5}$ $\Omega$-m as the temperature increases up to 50 K. Subsequently, the resistivity further decreases to $1\times10^{-5}$ $\Omega$-m at 300 K. These distinct decreases in resistivity separated by a plateau around 50 K suggest the presence of multiple gaps in $FeSb_2$. To analyze this behavior, the data below and above the plateau were fitted using the Arrhenius equation, $\rho = \rho_o exp\left(\frac{E_g}{2k_BT}\right)$, revealing the existence of two band gaps (supplemantary material, **Figure S5**). The estimated values for these band gaps were found to be in the range of 0.93-1.46 meV and 22-27 meV, respectively. It is noteworthy that these band gap values are smaller than those reported in previous literature.[16,20] At around 6.5 K, an intriguing observation is the clear manifestation of a metal-insulator transition (MIT) in pure $FeSb_2$ and Bi-substituted samples, indicating the presence of metallic states below this temperature. This phenomenon was initially identified by Xu et al., by employing angle-resolved photoemission spectroscopy (ARPES) measurements. Their research confirmed the presence of metallic surface states in the (010) plane of $FeSb_2$ single crystals.[22] These studies were conducted with the intention of unraveling the enigmatic source of the observed topological behavior in $FeSb_2$. Our research further supports the presence of metallic surface states in $FeSb_2$ below 6.5 K, as demonstrated by the intact MIT transition observed in $FeSb_2$ and iso-electronically Bi-substituted sample. However, this transition disappears in a Pb-substituted sample. Upon the substitution of Bi and Pb, the electrical resistivity increases to ~$12\times10^{-5}$ and ~$12\times10^{-4}$ Ω-m, respectively, in comparison to ~$9\times10^{-5}$ Ω-m for pure $FeSb_2$ at low temperatures. This increase in resistivity can likely be attributed to the induction of defects or additional grain boundaries (**Figure 4**), which act as scattering centers for the electrons traversing the material. Moreover, it is noteworthy that the plateau around 50 K fades out after the substitution, indicating a partial filling of the small transport gap.[16] This observation suggests a modification in the electronic properties of the material as a result of the introduced substitutions.

The pure $FeSb_2$ sample exhibits a significantly enhanced thermopower within the temperature range of 30-50 K, along with relatively low electrical resistivity in the same temperature range. This unique combination results in a maximum power factor (PF) value of ~1015 $\mu WK^{-2}m^{-1}$, occurring at approximately 42 K (**Figure 3c**). However, upon introducing Bi and Pb substitutions, the peak PF value decreases to ~587 and ~239 $\mu WK^{-2}m^{-1}$, respectively. Notably, the peak PF values for the Bi and Pb substituted samples not only decrease but also shift to lower temperatures of 35 and 40 K, respectively.

Subsequently, the thermal conductivity of all three samples was measured and is depicted in the **Figure 3d**. To determine the lattice thermal conductivity ($\kappa_{lat}$), the electronic thermal conductivity ($\kappa_e$) was subtracted from the total thermal conductivity. The value of $\kappa_e$ was estimated using the Wiedemann-Franz law, $k_e = \frac{L_0 T}{\rho}$ where $L_0$ is the Lorenz number ($2.44 \times 10^{-8}$ V$^2$K$^{-2}$) for metals,[24] $T$ is the temperature, and $\rho$ is the resistivity. The results clearly indicate that the contribution of $\kappa_e$ to the total thermal conductivity is negligible below 100 K, suggesting that variations in $\kappa(T)$ below 100 K are primarily governed by lattice vibrations or phonons. The highest $\kappa$ value observed was ~15 W/m-K for the pure FeSb$_2$ sample at approximately 57 K. However, when Bi and Pb were substituted into the sample, the $\kappa$ values decreased to ~10 and ~8.7 W/m-K, respectively, with a slight shift towards higher temperatures (around 65 K). This decrease can be attributed to increased phonon scattering resulting from the presence of additional grain boundaries in the substituted samples, as supported by electron backscattered diffraction (EBSD) analysis shown in **Figure 4.** In the case of the pure FeSb$_2$ sample, the average grain size ranged from 2-4 μm, providing the longest mean free path for lattice vibrations and consequently the highest thermal conductivity. Conversely, the average grain size decreased to 0.9 μm and 2 μm for the Bi and Pb substituted samples, respectively. Despite the anticipated lower $\kappa$ based on average grain size, the substantial reduction in $\kappa(T)$ for the latter sample can be attributed to the generation of crystal defects or substitutional disorder, leading to the presence of very small grains (0.2-0.3 μm) within the sample, as depicted in the **Figure 4c.**

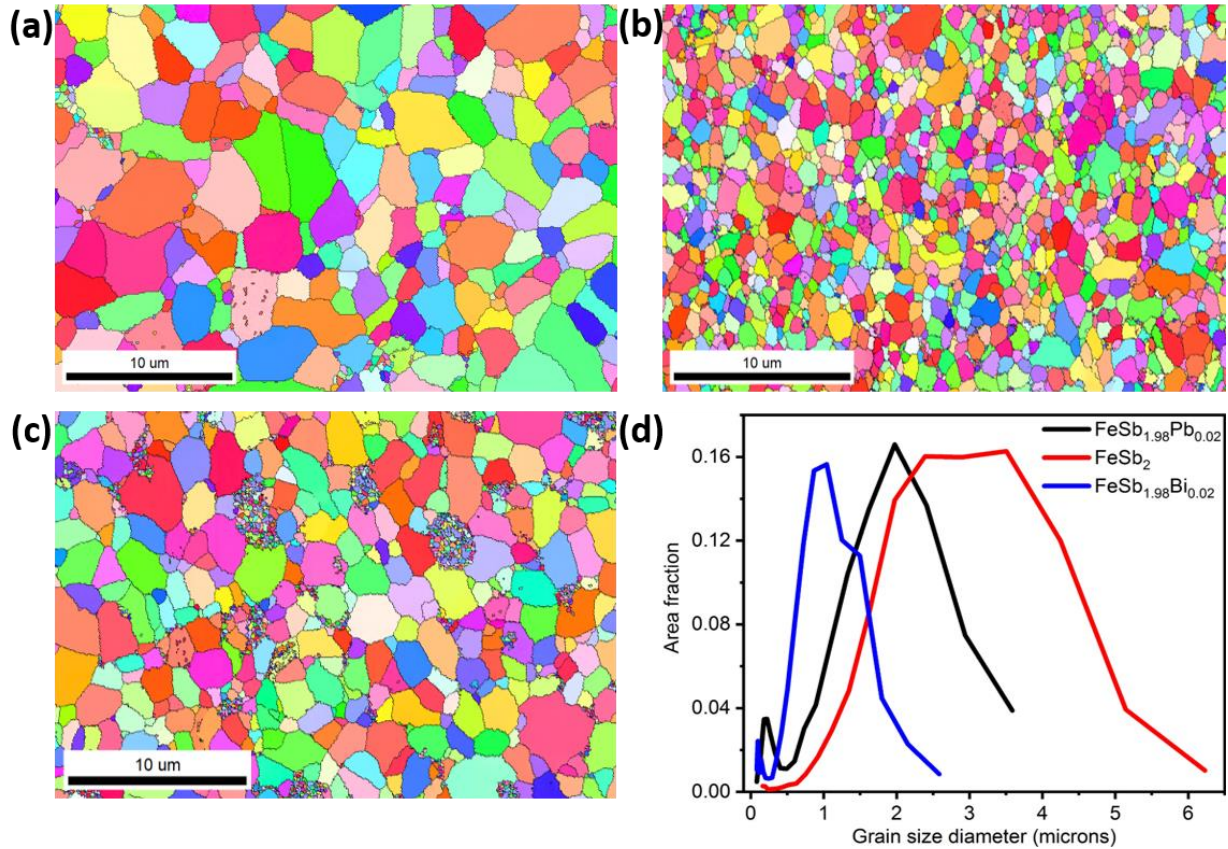

**Figure 4.** Electron back scattered diffraction (EBSD) images of spark plasma sintered samples (a) pure $FeSb_2$, (b) $FeSb_{1.98}Bi_{0.02}$, & (c) $FeSb_{1.98}Pb_{0.02}$ and their corresponding grain size distribution.

The Hall effect measurements $R_H(T)$ were conducted on all samples in a magnetic field range of -12 to 12 T at various temperatures, and the results are presented in the **Figure 5**. The $R_H$ values were obtained by extracting the slope of the linear plot of $R(\Omega)$ versus the applied magnetic field (T) within the range of -5 to 5 T (supplementary material, **Figure S6**). The choice of limiting the slope calculation to 5 T is because the substituted samples exhibited deviations from the linear behavior at higher fields, which falls outside the scope of this study. For all samples, $R_H(T)$ displays negative values below 80 K, indicating the dominance of electronic conductance. However, at higher temperatures, $R_H$ changes sign to positive for all samples, indicating the presence of holes as charge carriers (supplementary material, **Figure S7**). The significant increase in $|R_H(T)|$ as the temperature decreases clearly indicates the narrow-band gap characteristic of $FeSb_2$. In the pure $FeSb_2$ sample, a shoulder is observed around 40 K, corresponding to the plateau observed in $\rho(T)$. This shoulder is absent in both substituted samples, confirming the partial filling of the small energy transport gap. This disruption is also supported by the absence of a plateau in the temperature dependence of resistivity $\rho(T)$ in the substituted samples. A dip in the $|R_H(T)|$ value around 7 K is observed in both the pure $FeSb_2$ and $FeSb_{1.98}Bi_{0.02}$ samples, confirming the occurrence of a metal-insulator transition within this temperature range. The variation in the absolute values and positions of the $|R_H(T)|$ peaks among the three samples indicates the extreme sensitivity of the carrier concentration to slight substitutions in the $FeSb_2$ sample.

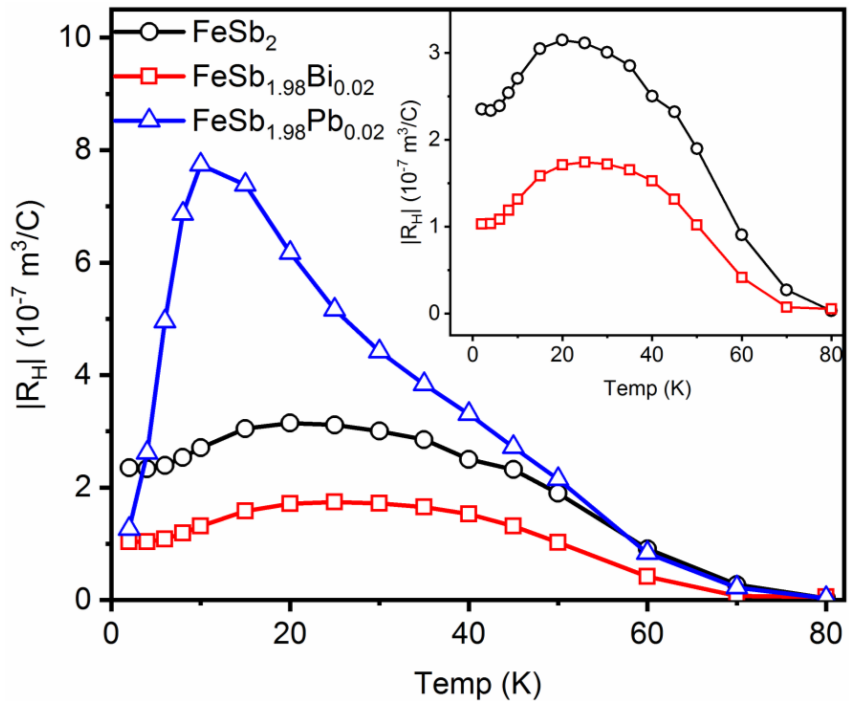

**Figure 5.** (a) Absolute values of the Hall coefficient $|R_H(T)|$ of $FeSb_2$, $FeSb_{1.98}Bi_{0.02}$, and $FeSb_{1.98}Pb_{0.02}$ samples.

Next, the figure of merit, $zT$, for all the samples is presented in **Figure S8** (supplementary material). The results clearly demonstrate that the substitution of heavy metals leads to a decrease in $zT$ value. While the substitution of Bi and Pb reduces the $\kappa$ value by up to 31 and 41 %, respectively, compared to pure $FeSb_2$. However, the reduction comes at the cost of degrading the Seebeck and electrical conductivity values, ultimately diminishing the $zT$ value. Notably, the substitution of Bi and Pb has shifted the $zT$ maximum to lower temperatures. If an effective approach can be developed to reduce $\kappa$ without compromising the Seebeck and electrical conductivity values, these substitutions would be crucial for achieving a high $zT$ value across the entire temperature range of 10 to 77 K.

## 3. CONCLUSIONS

In summary, we have successfully synthesized the narrow band-gap semiconductor $FeSb_2$, along with two substituted compounds, $FeSb_{1.98}Bi_{0.02}$ and $FeSb_{1.98}Pb_{0.02}$, and investigated their electrical and thermal transport properties. The unsubstituted $FeSb_2$ exhibits two distinct transport gaps, ranging from 0.93-1.46 meV and 22-27 meV, respectively. The substitution of these heavy elements met our expectations in terms of reducing thermal conductivity. Unfortunately, this does not lead to an improvement in the $zT$ value due to overcompensation caused by the reduced Seebeck coefficient and electrical conductivity. Notably, our synthesis method enabled us to shift the $zT$ maximum from the theoretically predicted 10 K to 45 K. This advancement opens up the possibility of extending the temperature range up to the boiling temperature of $N_2$ (77 K). More importantly, our investigation yielded compelling evidence that metallic surface states are present below 7 K in both $FeSb_2$ and the $FeSb_{1.98}Bi_{0.02}$ sample. This discovery definitively addresses a longstanding inquiry regarding the resemblance of $FeSb_2$ to 4$f$ topological Kondo insulators.

**Supplementary material section**

The supporting information summary includes experimental section, which provides a detailed account of the chemicals employed, the sample preparation technique, and the equipment utilized for characterization. **Figure S1** presents the X-ray diffraction (XRD) patterns of $FeSb_2$ with Bi/Pb substituted samples with varying concentrations of 1-5 %. **Figure S2** represents the refined XRD patterns of powder and SPS samples with refined lattice paramters displayed in **Table S1**. **Figure S3** displays the Williamson-Hall plot, which incorporates all relevant parameters outlined in **Table S2**. **Figure S4** showcases scanning electron microscopy (SEM) images of the as-synthesized powder samples. **Table S3** provides the elemental composition of all three samples after spark plasma sintering (SPS) treatment. The Arrhenius plot for $FeSb_2$ is illustrated in **Figure S5**, while **Figure S6** presents resistance-magnetic field plots. Furthermore, **Figure S7** explores the temperature dependence of charge carriers. Lastly, **Figure S8** presents the figure-of-merit (zT) values of all three samples at different temperatures.

**Acknowledgment**

We acknowledge the support of Institute Instrumentation Centre (IIC), IIT Roorkee for providing characterization facilities (XRD, SEM, PPMS, EBSD) and Department of Metallurgical and Materials Engineering, IIT Roorkee for providing spark plasma sintering (SPS) facility. D. Gujjar acknowledges the senior research fellowship (SRF) granted by University Grants Commission (UGC), India.

**Competing Interests**

The authors have no relevant financial or non-financial interests to disclose.

**Author Contributions**

**D. Gujjar:** Methodology, Investigation, Data Curation, Writing-Original Draft, Validation, Formal analysis, Visualization **S. Gujjar:** Synthesis & Data Curation **V. K. Malik:** Conceptualization, Writing-Review & Editing **Hem C. Kandpal:** Conceptualization, Writing-Review & Editing, Supervision, Project administration, Funding acquisition.

**Data availability statement**

The datasets generated or analysed during the current study will be available from the corresponding author upon reasonable request.